# Redefining Accountability: Navigating Legal Challenges of Participant Liability in DAOs

*Aneta Napieralska (http://orcid.org/0009-0004-0463-3646) (5238c665-86c0-47eb-b55c-18d220b8db42)*

*University Humanitas, Poland*

*Przemysław Kępczyński (http://orcid.org/0009-0006-8427-3162) (9fd76657-acf6-4f73-a6ea-ba063c711e7a)*

*University Humanitas, Poland*

## 1. Introduction

In the era of digitization and rapid technological progress, DAOs are gaining significance as innovative organizational forms that utilize blockchain technology to redefine traditional management and operational structures. These decentralized and autonomous organizational units base their operations on so-called smart contracts, which automate decision-making and execution processes, eliminating the need for intermediaries and enhancing operational efficiency[1]. This scholarly work aims to thoroughly investigate and define the organizational nature of DAOs and the implications of this understanding for the principles of legal liability of its participants.

The central thesis of this work posits that the proper definition of DAOs and the interpretation of smart contracts used within these organizations have a direct impact on the mechanisms for attributing liability to their members. The absence of clearly defined legal frameworks for DAOs, both nationally and internationally, complicates the application of traditional legal mechanisms to resolve liability issues. Given this regulatory gap, there is an urgent need to understand how current law can be interpreted or adapted to the unique features of DAOs.

This study aims to examine fundamental issues necessary for understanding liability in DAOs, focusing on the crucial issue of the definition and nature of the obligations that bind participants, especially role of smart contracts. Additionally, the work will focus on discussing the impact of the unique features of DAOs on the traditional understanding of liability and how the law might respond to ensure fair and effective regulations.

This publication serves as an introduction to a broader discussion of liability in DAO organizational structures. The authors base their analysis on the fundamental principles of this legal institution, which, with certain modifications, are common to most European legal systems. This shared heritage, rooted in ancient Rome, allows for general observations on this issue and highlights the need for further research[2].

These considerations are an attempt to lay the groundwork for future discussions on the regulation and narrativization of DAOs as full-fledged participants in the legal market, which is crucial for their stability and credibility in the global economic ecosystem.

## 2. Understanding legal liability in the context of Decentralized Autonomous Organization

The concept of liability in traditional organizations, characterized by well-defined structures, clear division of roles, and regulation under established laws, has been the subject of extensive research and forms the basis for diverse and sometimes conflicting approaches. Liability encompasses a multitude of factors that, when interpreted together, determine whether the preconditions for liability have been met, who is liable, and to what extent. Private law also recognizes circumstances that may exclude or modify liability, such as force majeure (vis major), a concept with roots in ancient Rome. Liability can manifest in various forms, including personal liability, entity liability, and collective liability, each with its own distinct characteristics[3].

---

[1] Savelyev, A.: Contract Law 2.0: 'Smart' Contracts as the Beginning of the End of Classic Contract Law. Information & Communications Technology Law 26(2), 116–134 (2017).

[2] Zimmermann, R.: The Law of Obligations: Roman Foundations of the Civilian Tradition. Oxford University Press, Oxford (1996)

[3] Zimmermann, R.: The New German Law of Obligations: Historical and Comparative Perspectives. Oxford University Press, Oxford (2005)

DAOs requires an introduction to basic civil law concepts (type of liability, source of liability, criteria for assigning liablility). Attributing liability necessitates the definition and identification of numerous factors. Primarily, it is essential to establish the source of liability, which in private law can be either contract or tort. Subsequently, the liable party must be identified, the harm must be defined, and a causal link must be established between the occurrence of the harm and the actions of the responsible party. Understanding the essence and challenges of determining each of these factors in the context of DAOs is, in the authors' view, a critical undertaking. Proper identification of these prerequisites will enable the lawful attribution of liability to a member of the autonomous organization.

### 2.1. The Choice of Liability Regime: Navigating the Complexities of Contractual and Tort Liability in DAOs

Legal liability (civil definition) refers to the obligation of an individual or group to remedy harm caused to others, based on civil law provisions, which can take the form of contractual liability (ex contractu) or tort liability (ex delicto)[4].

Contractual liability arises when a party to a contract fails to fulfill its contractual obligations, resulting in harm to the other party. It is characterized by improper performance or complete non-performance of the obligations arising from the contract. In the context of DAOs, the interpretation and execution of smart contracts, which are the legal form of obligations between DAO participants, are crucial. Adequately adapting the principles of imputing liability to encompass the unique characteristics of algorithmic contracts poses a significant challenge for legal scholars and practitioners.

Tort liability pertains to damages caused by actions unrelated directly to contracts, which violate the rights of others. It is liability for wrongful acts that lead to property or personal damage outside the scope of contractual agreements. In the context of DAOs, this could include, for example, data security breaches or improper management of the organization's resources by participants.

The selection of an appropriate liability regime will significantly impact the manner in which a claim for damages is pursued. As will be discussed in greater detail in the subsequent sections of this article, establishing contractual liability for members of a decentralized autonomous organization (DAO) presents a substantial challenge due to the algorithmic nature of the contract binding the participants[5].

Given the scope of this work, we will primarily focus on the contractual liability of DAO members, as it is directly linked to the interpretation and execution of smart contracts. However, it is crucial to acknowledge that tort liability may also arise in the context of DAOs, particularly in situations involving harm caused by the DAO's actions or omissions that are not directly related to contractual obligations.

### 2.2. Determining the Scope of "Damage" in DAOs

In the context of DAOs , a crucial task lies in defining the scope of what constitutes "damage" under civil law principles[6]. The concept of damage typically revolves around the notion of financial loss, raising the question of whether the loss of tokens, a drastic decline in their value, or other events leading to a "depletion" compared to the pre-DAO membership state can be considered damage.

It is important to note that civil law recognizes various types of damage. Damage can encompass both financial loss and lost profits. However, proving the existence of lost profits often presents a more complex challenge compared to demonstrating financial loss. The latter, as the term "demonstration" itself implies, requires more than just a subjective perception of harm. In such proceedings, it is essential to prove and substantiate the occurrence of damage and to precisely determine its extent[7]. This necessitates sound arguments and the collection of evidence that can convince a court of the existence of damage during the proceedings.

Considering the relatively novel nature of DAO operations globally, it is conceivable that courts may be reluctant to recognize financial losses caused by the actions of DAOs or their members as civil law damages. One argument could be that tokens used by these organizations do not constitute property rights, and therefore, their value loss might not be recognized as damage under current legal frameworks[8]. In the absence of one of the prerequisites for liability, further proceedings in such cases could lack legal grounds. This hypothetical situation highlights the ongoing debate surrounding token definitions and the regulation of the crypto asset market, even with the introduction of the MiCA Regulation. These considerations are based on current legal discussions and the evolving nature of digital assets in the legal landscape. The key element is determining the moment of occurrence of the damage and the associated direct detriment, which is fundamental for attributing liability.

---

[4] Treitel, G. H.: The Law of Contract. 14th edn. Sweet & Maxwell, London (2015)
[5] De Filippi, P., Wright, A.: Blockchain and the Law: The Rule of Code. Harvard University Press, Cambridge, MA (2018)
[6] Zimmermann, R.: The Law of Obligations: Roman Foundations of the Civilian Tradition. Oxford University Press, Oxford (1996).
[7] H.A. Cousy, Wrongfulness in Belgian Tort Law, w: H. Koziol (red.), Unification of tort law: Wrongfulness, The Hague–London–Boston 1998, s. 33–34
[8] F. Guillaume, "Decentralized Autonomous Organizations (DAOs) Before State Courts. How Can Private International Law Keep Up With Global Digital Entities?", in: Perestrelo de Oliveira/Garcia (eds), DAO Regulation: Principles and Perspectives for the Future, 2023.

### 2.3. Causation in DAOs : Navigating a Complex Landscape

Causation, the cornerstone of establishing liability in both contract and tort law, presents a significant challenge in the context of DAOs. Here, liability hinges on demonstrating an adequate causal relationship between the conduct or omissions of the DAO (or its members) and the harm suffered by another party[9]. Traditional causation analysis, often revolving around a "but-for" test and considerations of foreseeability, proves insufficient when applied to the intricate structure and operations of DAOs[10].

The decentralized nature of DAOs, with their reliance on algorithmic contracts and potential for involvement by multiple actors, complicates the task of identifying the causal chain of events. Unlike traditional entities with centralized decision-making, DAOs may involve a complex interplay between smart contracts, user actions, and potentially unforeseen circumstances. This distributed decision-making process creates a challenge in pinpointing the source of the causal act[11].

**Smart Contracts: A Double-Edged Sword**

Smart contracts, the self-executing agreements that govern DAO operations, introduce a double-edged sword when it comes to causation. On the one hand, their automated nature raises questions about the locus of causation – who or what is truly responsible for the actions dictated by the code? This ambiguity may potentially shield individuals or entities associated with the DAO from direct liability claims[12].

On the other hand, smart contracts can serve as a valuable tool for establishing causation. Their immutable and transparent nature allows for a clear record of the actions and decisions that led to the harm. By analyzing the code and its execution history, one can potentially trace the causal chain of events. However, effectively utilizing smart contracts as evidence necessitates a deep understanding of blockchain technology and the ability to extract relevant data for forensic analysis.

### 2.4. Identification of the Liable Party- importance of structure

In the context of DAOs, determining who is liable for damages can be challenging due to the anonymity of participants and the decentralized structure of management.

**Determining Liability in Hybrid DAOs and Informal DAOs**

In Hybrid DAOs, where the organization is operated by legal entities, the identification of individuals subject to liability principles adheres to the norms of generally applicable law. Typically, individuals responsible for the DAO's actions will be identical to the boards or managers of the entity leading the decentralized autonomous organization. Alternatively, the entity itself may bear liability, possessing legal personality that allows claims to be directed towards it directly. Additionally, a scenario exists where the entity bears primary responsibility, but managers may also incur personal liability if participant harm can be attributed to their actions or omissions. This determination will usually be resolved by applying the relevant norms governing the entity's operation.

In the event of harm to both participants and external parties, it appears that these individuals will be primarily liable. The authors posit that determining individuals responsible for the actions of DAO those not organized according to generally applicable norms, is more complex. While legal entities can be held liable relatively easily, attributing liability to an informal DAO itself presents challenges. In such cases, claims would need to be directed towards each member individually, with satisfaction by one member releasing the others from responsibility[13]. This scenario would necessitate the application of the well-established regime of joint and several liability within European legal systems.

The identification of specific individuals responsible within an informal structure hinge on the provisions of the smart contract. Does the DAO structure designate specific roles for individual members, granting them additional rights or obligations? The fulfillment or non-fulfillment of these obligations can form the basis of liability for these specific members towards other participants, most often falling under the contractual liability construct. Individual members, bound by the obligation established with other members, would be obligated to perform specific actions. Consequently, a failure to fulfill such obligations could give rise to liability. Therefore, a crucial aspect in this context is the precision and correctness of the agreement concluded between members, taking the form of an algorithmic contract.

---

[9] Hart, H.L.A., Honoré, A.M.: Causation in the Law. 2nd edn. Clarendon Press, Oxford (1985).
[10] P. Machnikowski, w: System PrPryw, t. 6, 2009, s. 434–438, Nb 38–40
[11] Wright, A., De Filippi, P.: Decentralized Blockchain Technology and the Rise of Lex Cryptographia. SSRN Electronic Journal (2015)
[12] Fenwick, M., Kaal, W. A., Vermeulen, E. P. M.: Regulation Tomorrow: What Happens When Technology Is Faster Than the Law? American University Business Law Review 6(3), 561–594 (2017).
[13] Vogenauer, S., Weatherill, S.: The Harmonisation of European Contract Law: Implications for European Private Laws, Business and Legal Practice. Hart Publishing, Oxford (2006).

**Structured DAO**

In the context of DAOs that implement a structured governance framework, establishing a causal link between the actions of authorized members and potential harm to participants is potentially facilitated. This is due to the clear identification of liable entities and the inherent traceability of actions enabled by smart contracts.

Structured DAOs often establish a hierarchy of membership with varying levels of authority and decision-making power. This clearly defines the individuals or entities that can be held accountable for their actions within the organization. Unlike traditional corporate structures, where the attribution of liability may be more complex due to the involvement of multiple actors and decision-making layers, structured DAOs offer a more streamlined approach to identifying liable parties.

The self-executing nature of smart contracts, the backbone of DAO operations, provides an inherent mechanism for tracing actions and establishing a clear record of decision-making processes. Unlike traditional forms of governance that may rely on documentation and human testimony, smart contracts offer an immutable and transparent record of the actions taken and the decisions made. This facilitates the reconstruction of events and the identification of specific actions that potentially led to harm.

In structured DAOs, proving an adequate causal link between the actions of authorized members and the harm suffered by participants often revolves around the interpretation of smart contract provisions. This involves analyzing the code to determine which individuals or entities were granted the authority to take specific actions and whether those actions contributed to the occurrence of the harm.

Beyond traditional organizational roles and individuals potentially liable to other members, it is also necessary to consider the group responsible for the DAO's technical aspects. Individuals or entities providing services to create the appropriate technical infrastructure for DAO operation and management may be liable in the event of improper smart contract code functioning. The legal basis (source of obligation) upon which they created the code would also serve as the foundation for contractual liability towards individual members.

**Considerations in Anonymous Membership Situations**

Finally, it is crucial to emphasize that the aforementioned considerations may be significantly modified in situations where membership in the organization is anonymous. In such cases, while technological tools can be used to identify specific individuals who have acted within the structure, this process is not only expensive and complex but also inaccessible to the average person. The greater likelihood of obtaining such information arises in the context of criminal liability, where state authorities possess appropriate evidentiary tools.

### 2.5. "Autonomous" and "Regulated" DAO- two different approach

Due to the extensive nature of this article, the authors focus on a general discussion of the most important issues related to the concept of liability in DAOs. Considering the foregoing, it is undoubtedly appropriate to consider and divide DAOs into those that operate within a known and regulated legal form- "Regulated DAO" and those that function as a tool for enabling a community to operate- "Autonomous DAO".

**"Regulated" DAOs.** In the case of organized forms, defining the principles and scope of liability of its members will be based on the general principles implied for the specific form of the organization's activity[14]. At the same time, these principles will have to undergo some modifications due to the nature and manner in which such organizations operate. It should be noted here that in traditional organizational forms, the principles of so-called corporate governance are one of the most important areas of the organization's activity. It is this that determines how the entity is managed, how decisions are made, how these decisions can be challenged, what are the relationships and powers of the participants or members of such an entity, or how ownership conflicts are resolved. In the case of DAO that operate using an official legal form, it should be assumed that traditional corporate governance principles should also be reflected in the provisions contained in the smart contract. As a rule, European legal systems, such as the German, French, or Polish legal systems, in their cognitive norms concerning the functioning and principles of liability in legal entities, define the principles of liability in detail. Therefore, the content of smart contracts that form the basis of the activity of DAOs should be consistent in this respect with generally applicable regulations.

The question arises when the provisions contained in the smart contract are not consistent with the principles of corporate governance defined as mandatory provisions, which constitute the regulatory minimum imposed by a given jurisdiction. In this situation, in place of provisions that are inconsistent with cognitive regulations, general principles will most often come into play. This will also have a significant impact on the adopted principles of liability. In turn, inconsistencies between the smart contract - its provisions - and the general norms concerning corporate governance may give rise to liability for the persons who created the smart contract norms, implemented it, and were obliged to manage the organization and represent it externally.

---

[14] Kraakman, R., et al.: The Anatomy of Corporate Law: A Comparative and Functional Approach. 3rd edn. Oxford University Press, Oxford (2017).

**Hybrid DAOs.** One can also imagine a situation where the organizational structure of a decentralized autonomous organization, from the perspective of the organization of a legal entity, will be characterized by the fact that in reality the decentralized autonomous organization will function as a form of activity of an official legal entity (two organization- official and unofficial)[15]. It is assumed that these are so-called hybrid DAOs or DAOs with a legal wrapper. Such structures are most often created by investment funds, which on the one hand want to ensure a certain traditional form of operation that enables interaction with external entities, and on the other hand, they use DAO mechanisms for investing and also raising capital. An example of such a structure is Moloch DAO, operating under the auspices of a Foundation, which organized and enables the current operation of the DAO. In principle, this vehicle plays a kind of Venture Capital role, directed at a wider audience around the world. In this type of structure, in the authors' opinion, the majority of the responsibility will rest with the entity that organized the decentralized organization, and the DAO members should be treated rather as service providers and not organizers, due to the significant limitation of decision-making possibilities regarding the organization and form of operation of the vehicle.

While in the case of DAOs that are created by officially functioning entities - companies, associations, etc., defining clear principles of liability and the persons who will be subject to such liability seems to be a straightforward action. At the same time, it should be pointed out that despite the existence of clear rules that will define who and how is responsible for the functioning of the organization, due to the complexity of the institution of civil liability, even attributing liability to anyone, defining the injured parties, or determining the damage may already be more complicated due to the innovative nature of DAO operation.

### 2.6. Liability in "Autonomous" DAOs

Transitioning to autonomous DAOs, the intricacies of determining liability intensify. In principle, directly implicating liability rules derived from private or corporate law onto an existing official structure is problematic. After all, the fundamental tenet of establishing DAOs was to emancipate these organizations and their frameworks from universally applicable norms, effectively excluding the jurisdiction of ordinary courts.

In the authors' view, such a construct represents a form of social activity undertaken by a group of individuals utilizing technology for a shared objective[16]. The foundation for collaboration among participants in a decentralized autonomous organization will, in this case, be an agreement between DAO participants, albeit one formulated in an unconventional format - an algorithmic contract.

The very sources of liability - principles in this instance - must be sought within the smart contract's content. The scope of liability, the definition of participant roles within this organization, and the mechanisms for asserting their rights will all be determined by the contract's content. While on the one hand, the autonomy of this type of organization may manifest itself precisely in the fact that all the rules governing the functioning of this type of organization will be contained in the smart contract, on the other hand, the provisions of this agreement will also be subject to interpretation on the basis of generally applicable regulations.

In the authors' assessment, the issue of liability in DAOs will be inextricably linked to the interpretation and substance of smart contracts. Conversely, the interpretation of the agreement's provisions binding the organization's participants will typically occur within the framework of generally applicable civil law regulations. In this scenario, we can unequivocally assert that participants in a decentralized autonomous organization will be subject to both contractual liability and liability stemming from tort.

It is noteworthy that each member of the organization would be entitled to a claim for damages individually. Since a DAO lacking a legal and recognized form of operation does not possess legal personality, it cannot be a subject of rights and obligations. This, in turn, can significantly complicate the process of claiming damages and proving harm, particularly in organizations with thousands of members. Such fragmentation would practically render effective claim enforcement impossible.

With regards to identifying those responsible for harm caused by the "organization" through its actions towards individuals not involved in the project (tort liability), the authors posit that each member will be liable. Here, the key factor in assigning ultimate liability will be demonstrating that a specific user's action or omission led to the occurrence of the damage.

---

[15] Zaczyk M. : The concept of a decentralized autonomous organization as an innovative organizational structure, Silesian University of Technology Publishing House, 2023
[16] R. Braakman, M. J. Follows and S. W. Chisholm, "Metabolic evolution and the self-organization of ecosystems", Proc. Nat. Acad. Sci. USA, vol. 114, no. 15, pp. E3091-E3100, 2017

# Overview of Liability in the Context of Smart Contracts in DAOs

Traditional contractual liability principles, deeply rooted in the actions and intentions of parties, face challenges when applied to DAOs and smart contracts. The self-executing nature of smart contracts, the limited control parties have over their execution, and the predefined conditions for liability raise fundamental questions about determining breach, causation, and damages[17].

Authors research indicates that smart contract forms the basis of contractual liability for participants in DAOs. The interpretation of these contracts is crucial for determining the rights and obligations of community members. Consequently, the proper functioning of the code directly influences the ability to assign liability. In traditional contracts, the focus is on whether the obligations have been fulfilled in accordance with the contract. However, in the case of smart contracts, the fulfillment of a contractual obligation undergoes a modification of meaning: by joining a smart contract, a participant automatically agrees to its stipulations, which is comparable to fulfillment of an obligation by a third party.

### 2.7. Smart Contracts as the Foundation of Legal Relationships in DAOs

In decentralized autonomous organizations (DAOs), smart contracts function analogously to contracts in traditional civil law, serving not only as a form of establishing obligations but also as a means of their existence and performance. As a set of rules encoded in computer code, smart contracts precisely define the rights and obligations of DAO participants, similar to how traditional contracts define the rights and obligations of parties. In this sense, a smart contract can be considered a digital manifestation of the parties' mutual consent, aiming to establish a legal relationship, which is the essence of a contract according to civil law principles (e.g., Article 66 of the Polish Civil Code).

The principle of freedom of contract, a cornerstone of civil law, also applies to DAOs. DAO participants have the freedom to shape the content of smart contracts, allowing them to flexibly adapt the organization's operating principles to their needs and goals. For example, DAOs can define decision-making processes, profit-sharing mechanisms, and even dispute resolution procedures within their smart contracts.

Smart contracts fit into the natural evolution of forms for establishing obligations, reflecting technological and social progress. From ancient oral forms, such as *stipulatio* in Roman law, through written forms, to modern electronic forms like email or electronic signatures, each era has introduced new solutions adapted to the current realities. Smart contracts, as a digital form, represent the next stage in this evolution, enabling the establishment of obligations in an automated, transparent, and secure manner.

In the context of DAOs, smart contracts function as both framework agreements, defining the general operating principles of the organization, and detailed agreements, regulating specific transactions and interactions among participants. Like traditional civil law, obligations arising from smart contracts can be classified as *essentialia negotii* (essential elements of the contract), *naturalia negotii* (customary elements), or *accidentalia negotii* (additional elements agreed upon by the parties).

Smart contracts, as digital contracts, form the basis of contractual liability for DAO members. Violating the rules set out in a smart contract can result in contractual liability towards other DAO members or the entire organization. In practice, this means that DAO members can be held accountable for failing to fulfill or improperly fulfilling their obligations arising from the smart contract, which may lead to the necessity of repairing damage or paying compensation.

However, the decentralized nature of DAOs and the specific characteristics of smart contracts pose certain challenges in defining and enforcing liability. Anonymity of members, lack of clear legal frameworks, and difficulties in identifying and quantifying damages are just some of the issues faced by DAOs. Nevertheless, as a digital contract, a smart contract serves as a starting point for analyzing liability in DAOs and allows for the application of many principles and institutions known from traditional civil law, such as the principle of fault, adequate causation, and the duty to repair damage.

Yet, in the context of smart contracts, the question arises whether improper functioning of the contract directly burdens a community member, or whether they can invoke circumstances that exempt them from liability. The operation of a smart contract can be compared to the formulaic process in ancient Rome, where the praetor was limited by whether the parties properly presented the required formulas. Similarly, a smart contract does not interpret the factual state but merely checks whether predefined conditions have been met.

---

[17] Mahoney P. "Contract or Concession? An Essay on the History of Corporate Law" (2000) 34(2) Georgia Law Review 873 at 880 and Alexander Fallis "Evolution of British Business Forms: A Historical Approach" (2017) <www.icaew.com/-/media/corporate/files/technical/ethics/evolution-of-british-business-forms.ashx?la=en>.
70 See generally, Ron Harris "The Private Origins of the Private Company: Britain 1862-1907" (2013) 33 Oxford Journal of Legal Studies 339.

Irregularities in the delivery or processing of data, which underpin the decisions of a smart contract, can significantly affect its functioning. For example, improperly programmed smart contracts may operate in a way not anticipated by participants, and deliberate data manipulation can lead to specific contract behaviors. In the case of errors, responsibility can be attributed to those who improperly prepared the contract, whereas in the case of manipulation, liability may be treated as a tort.

In addressing this issue, the use of a dual obligation construction and conducting tests before launching a smart contract can better understand and interpret the intentions of participants and the contract's compliance with the intentions of the parties. This approach allows for more flexible and secure management of obligations within DAOs, enhancing legal certainty and reducing the risk of conflicts.

### 2.8. Code is law vs Participant Will

In the light of the example of The DAO, where a participant, despite the intentions of other members, amassed a significant portion of the organization's funds, claiming that their actions were entirely consistent with the stipulations of the smart contract—implying that the contract would not have allowed such actions if they were not permissible—it becomes crucial to consider whether employing a dual-obligation structure and directing interpretations based on the traditional obligations formed through the exchange of consents would be a safer approach. This case underscores the importance of interpreting declarations of intent and how crucial this is in determining liability. The issues related to oracles and potential errors in the operation of the code also need to be extensively discussed. Furthermore, while smart contracts are responsible for organizing the operations of DAOs, this may manifest in such a way that within the rules encoded in the code, individual members of the organization are authorized to perform their own actions. A smart contract is akin to the constitution of the organization, or using terminology closer to capital market law, it acts as the statute of the organization. However, unlike traditional companies where the execution of the statute and the verification of actions taken in accordance with the statute are subject to interpretation, in the case of a smart contract, participants are authorized and indeed able to carry out such operations and actions that the smart contract enables[18].

Here lies a crucial piece of information vital for the issue of liability: actions consistent with the content of the smart contract, which yet are not consistent with the will of the parties that accompanied the agreements regarding the content of the smart contract, might be considered as improper performance of the obligation. In this case, it would be appropriate to refer to the thesis of the dual nature of the obligation. It may indeed happen, as in the previously mentioned example of The DAO, that the participant's actions are consistent with the code but not with the agreement valued between the participants.

Given the episode involving The DAO, where a participant, contrary to the will of other members, accumulated a significant portion of the organization's resources, justifying that such actions were fully compliant with the smart contract's provisions—suggesting that the contract would not permit such actions if they were not allowed—raises a critical need to consider whether adopting a dual-obligation framework and guiding interpretations based on traditional obligations formed through mutual consent might be a safer approach. This scenario highlights the significance of interpreting will statements and their importance in determining liability. Furthermore, the issues surrounding oracles and potential code operational errors also require in-depth discussion.

Moreover, while smart contracts govern the operations of DAOs, this can manifest such that within the rules encoded in the code, individual organization members are empowered to undertake their own actions. In this respect, a smart contract acts somewhat as the constitution of the organization, or using terms familiar to capital market law, it serves as the organization's statute. Unlike traditional companies where the execution of the statute and the assessment of compliance with the statute are subject to interpretation, in the case of a smart contract, participants are authorized and, in fact, able to perform such operations and actions as the smart contract permits. This brings us to a crucial piece of information critical for understanding liability: actions that are consistent with the content of the smart contract yet do not align with the intentions of the parties involved in forming the contract might be regarded as improper fulfillment of the obligation. In such instances, referring to the concept of the dual nature of the obligation is pertinent. As illustrated by the earlier example of The DAO, a participant's actions, although consistent with the code, may not conform to the mutual agreement among participants.

Interpretation of the contractual obligations intended by the participants. This necessity is especially pronounced in environments governed by smart contracts, where the explicit programming may not fully capture the subtleties of human intentions and consensual agreements.

---

[18] Raskin, M.: The Law and Legality of Smart Contracts. Georgetown Law Technology Review 1(2), 305–341 (2017).

### 2.9. Importance of Testing and Interpreting Declarations of Intent

One potentially effective strategy, considering the dual-obligation concept, would be to implement a trial or testing phase for the smart contract before its final deployment. During this phase, the smart contract could be scrutinized under conditions that simulate real operational environments to assess whether its functionalities align with the agreed-upon intentions of the DAO members. Such testing could unveil discrepancies between the programmed behaviors and the participants' expectations, allowing for necessary adjustments before the contract is irrevocably executed.

Furthermore, a thorough examination of the declarations of intent that accompany the formation of the smart contract is crucial. This analysis helps ensure that the actions facilitated by the smart contract are not just mechanically correct but are also in true accordance with the collective will of the organization's members. This is vital because even actions that are technically in compliance with the smart contract's terms can sometimes diverge from the original consensual agreements among participants.

### 2.10. Addressing Potential Code Errors and Oracle Issues

The discussion also extends to the operational reliability of smart contracts, particularly concerning oracles and data feeds that inform contract behaviors. Errors in these external inputs can lead to significant misinterpretations and unintended actions by the smart contract. Addressing these concerns involves not only technical safeguards but also legal provisions that anticipate and mitigate potential failures in data accuracy and application.

It is essential to acknowledge that while a smart contract may act as the "constitution" of a DAO, the literal interpretation of its code may not always reflect the true spirit of the agreement among participants. Thus, if a participant's actions, although consistent with the code, violate the underlying contractual consensus, these actions should be evaluated against the dual nature of obligations. As demonstrated by The DAO incident, actions that adhere to the technical specifications of the smart contract but contravene the fundamental agreements among members must be scrutinized for their legality and fairness.

This underscores the need for robust frameworks that enable accurate interpretation of contractual obligations as intended by the participants. This requirement becomes particularly crucial in environments governed by smart contracts, where explicit programming may not capture all nuances of human intentions and consensual agreements.

**Importance of Testing and Interpreting Intent.** Implementing a testing phase for the smart contract before its final deployment could prove effective. During this phase, the contract could be evaluated in simulated real-world conditions to ensure its functionalities align with the DAO members' agreed intentions. Such testing would identify any discrepancies between programmed behaviors and participants' expectations, allowing for adjustments before irreversible execution. Moreover, it is crucial to thoroughly examine the declarations of intent accompanying the smart contract's formation. This ensures that actions enabled by the contract are not only mechanically correct but also genuinely reflect the collective will of the organization's members. This is vital because actions that technically comply with the smart contract's terms might still diverge from the original agreements among participants.

**Addressing Potential Errors in Code and Oracle Issues.** Oracles bridge the real-world data with the blockchain and are pivotal in triggering contract stipulations based on the data they feed into the system. However, this dependency introduces a significant risk factor: if the data sourced by oracles is inaccurate, manipulated, or delayed, it can lead to erroneous contract executions that may not truly reflect the agreed intentions of the parties involved. Incorporating specific clauses in the smart contract that address potential data inaccuracies and their consequences can help in predefining liability and enforcement measures. Legal provisions could include terms that allow for contract suspension or reversal in cases where data integrity issues significantly impact contract outcomes.

**Legal and Liability Concerns Arising from Data Manipulation.** From a legal perspective, the integrity of data inputs is not just a technical issue but a substantial liability concern. If a smart contract automatically executes a transaction based on faulty or manipulated data, determining who is at fault—and thus who is liable—can be complex. For instance, if data manipulation leads to financial loss for a party, the question arises whether the liability rests with the oracle provider, the data source, or the developers of the smart contract for not adequately vetting the data inputs.

Acknowledging that while a smart contract may serve as the "constitution" of a DAO, the literal interpretation of its code might not always embody the true spirit of the agreement among participants is essential. If actions, although consistent with the code, violate the underlying contractual consensus, they should be assessed against the dual nature of obligations. As seen in The DAO incident, actions conforming to the technical specifications of the smart contract but violating fundamental participant agreements must be scrutinized for their legality and fairness. To ensure the integrity of smart contracts and align them more closely with the true intentions of the DAO participants, a collaborative approach to contract creation and revision is recommended. This process would involve all stakeholders in the periodic review and updating of the smart contract terms to reflect changes in the organization's goals and the external environment. This collaborative revision process not only enhances transparency but also builds trust among members, reducing the likelihood of disputes.

### 2.11. The Two-Agreement Paradigm: A Nuanced Approach

To address these complexities and reconcile traditional contractual liability principles with the unique characteristics of DAOs and smart contracts, this analysis proposes a two-agreement paradigm[19] as a conceptual framework for understanding contractual liability in this context:

**Pre-Smart Contract Agreement.** This initial agreement arises through consensus among prospective DAO participants. It entails an obligation to collaborate in defining the smart contract's content and to subsequently join the DAO. This agreement establishes the foundation for the DAO's governance structure and the expectations of its members. The pre-smart contract agreement serves as a crucial foundation for establishing the legal relationships among participants and defining the scope of their obligations. It provides a framework for addressing potential breaches of pre-contractual duties, such as failing to collaborate in good faith or adhering to the agreed-upon terms for smart contract creation.

**Smart Contract Agreement**. Upon joining the DAO, participants enter into a second agreement governed by the smart contract itself. This agreement outlines their rights, obligations, and potential liabilities within the DAO structure. The smart contract serves as the operative framework for enforcing contractual obligations and determining liability. The smart contract agreement, with its self-executing nature and predefined conditions for liability, represents a departure from traditional contractual liability principles. However, it provides a mechanism for establishing clear expectations and ensuring that participants adhere to the agreed-upon terms of governance and collaboration.

The two-agreement paradigm offers a framework for understanding the formation, performance, and interpretation of obligations in DAOs:

**Formation of Obligations.** Pre-Smart Contract Agreement: The pre-smart contract agreement establishes the initial obligations among prospective DAO participants, including the duty to collaborate in creating the smart contract and adhering to its terms.

Smart Contract Agreement: Upon joining the DAO, participants enter into a second agreement governed by the smart contract, which outlines their specific rights, obligations, and potential liabilities.

**Performance of Obligations.** Pre-Smart Contract Agreement: Performance of obligations under the pre-smart contract agreement involves fulfilling the agreed-upon duties, such as collaborating in good faith and adhering to the terms for smart contract creation.

Smart Contract Agreement: Performance of obligations under the smart contract agreement involves fulfilling the specific duties outlined in the contract, such as contributing to the DAO's governance, adhering to its rules, and fulfilling financial commitments.

The concept of dual obligations in DAOs offers a framework for understanding the interplay between contractual liability and smart contracts. The first layer of obligations, represented by the pre-contractual consensus, allows for a degree of traditional interpretation, enabling participants to negotiate and express their intentions regarding the smart contract's operation. However, the second layer, which involves the execution of the smart contract, introduces an element of automation, where the obligations are fulfilled based on the pre-programmed instructions.

## Summary

Discussing these principles is essential to understand how traditional liability mechanisms are adapted or can be adapted to the specifics of DAO operations. This consideration of the foundations of liability in the context of DAOs is crucial for understanding how legal frameworks can be applied or need to be modified to adequately manage risk and protect the interests of both DAO participants and external parties. In traditional organizations, the analysis of liability must consider the well-established principles of liability, recognizing the diverse approaches and factors that influence liability determination. The unique characteristics of DAOs, such as their decentralized nature, anonymous participation, and automated decision-making, introduce new complexities to the assessment of liability. Understanding the legal implications of smart contracts, the self-executing agreements that govern DAO operations, is essential for assessing liability in DAOs. The governance structures and decision-making processes within DAOs can influence liability attribution. Identifying who has authority and control over DAO actions is crucial for assigning responsibility. Comparing and contrasting liability

---

[19] Napieralska A., Kępczyński P.: Smart Contracts and Web 3: From Automated Transactions to DAOs, Concepts, Technologies, Challenges, and the Future of Web 3, IGI Global, 2023

frameworks across different jurisdictions can provide valuable insights into potential approaches to addressing liability in DAOs.[20] The analysis of liability in DAOs is a complex and evolving area that requires careful consideration of both established legal principles and the unique characteristics of these decentralized organizations. Further research is needed to address the specific challenges and develop comprehensive frameworks for assessing and assigning liability in DAOs.

The focus of this paper is on civil liability (contractual and tort), while acknowledging that criminal liability in DAOs is an equally important and complex issue, particularly considering the potential anonymity of members, which significantly elevates the evidentiary burden. It is also worth noting that there is often a "mixing" of liability regimes when delict liability meets fraud crime or other financial crime. Finally, it is worth noting that in the context of DAOs whose purpose is the collection and investment of capital, in addition to the previously mentioned sources of liability, there may also arise criminal liability related to the issuance and trading of financial instruments or organizing so-called public fundraisers without the appropriate permissions[21]. Additionally, failure to meet the requirements imposed by the national regulator may also give rise to certain administrative liabilities. In summary, the governance of DAOs can engender a multiplicity of liability regimes. These range from contractual liabilities to criminal and administrative responsibilities. The specific delineation of the liability framework is contingent upon several factors: the operational modality of the DAO, its legal constitution, its intended objectives, the jurisdiction within which it operates, and, critically, the nature of the obligations that unite its participants.

---

[20] F. Guillaume, S. Riva, "Blockchain Dispute Resolution for Decentralized Autonomous Organizations", in: Bonomi/Lehmann (eds), Blockchain and Private International Law, Leyde 2023
[21] Jeremy I Senderowicz, K Susan Grafton, Timothy Spangler, Kristopher D Brown and Andrew J Schaffer "SEC Focuses on Initial Coin Offerings: Tokens may be Securities under Federal Securities Laws" (2018) 19 Journal of Investment Compliance 10.